\documentclass[a4paper,11pt]{article}
\usepackage{amsmath,amssymb,amsthm,amscd,a4wide}
\usepackage{epsfig,subfigure,graphics,graphicx}
\usepackage{color,xcolor}

\usepackage{kotex}

\usepackage{epstopdf}
\usepackage{booktabs}
\usepackage{ifpdf}

\usepackage{cite}
\usepackage[]{natbib}
\usepackage{natbib}
\usepackage{varioref,xr-hyper,hyperref}

\hypersetup{colorlinks,linkcolor=blue,citecolor=blue,filecolor=blue,urlcolor=blue}

\newtheorem{thm}{Theorem}[section]

\numberwithin{equation}{section}

\begin{document}
	
\title{\LARGE Income Disaster, Role of Income Support, and Optimal Retirement\footnote{Tae Ung Gang was supported by the National Research Foundation of Korea (NRF) under contract No. RS-2025-00513609, No. RS-2023-00237770, and Yong Hyun Shin was supported by the NRF under contract No. RS-2023-00240574.}}
	
\author{Tae Ung Gang\footnote{The Research Center for Natural Sciences, Korea Advanced Institute of Science and Technology (KAIST), Daejeon 34141, Republic of Korea.\;E-mail: {\tt gangtaeung@kaist.ac.kr}}\and
					\and
		Seyoung Park\footnote{Nottingham University Business School, University of Nottingham, UK.\;E-mail: {\tt seyoung.park@nottingham.ac.uk}}\and
					\and
		Yong Hyun Shin\footnote{Department of Mathematics \& Research Institute of Natural Sciences, Sookmyung Women's University, Seoul 04310, Republic of Korea. Corresponding author.\;E-mail: {\tt yhshin@sookmyung.ac.kr}
}}

\date{\today}
	
\maketitle \pagestyle{plain} \pagenumbering{arabic}

\thispagestyle{empty}


\abstract{This paper investigates the interactions among consumption/savings, investment, and retirement choices with income disaster. We consider low-income people who are exposed to income disaster so that they retire involuntarily when income disaster occurs. The government provides extra income support to low-income retirees who suffer from significant income gaps. We demonstrate that the decision to enter retirement in the event of income disaster depends crucially on the level of income support. In particular, we quantitatively identify a certain income support level below which the optimal decision is to delay retirement. This implies that availability of the government's extra income support can be particularly important for the low-income people to achieve optimal retirement with income disaster.}

\bigskip
	



\clearpage
\newpage
\setcounter{page}{1}


\section{Introduction}

\begin{quote}
Extra support could help low-income people approaching retirement age when the state pension is increased from 66 to 67 between 2026 and 2028. A new report from the Institute for Fiscal Studies, published as part of The Pensions Review, said targeted support could reduce the disproportionate impact the state pension age change will have on low-income people. (Financial Times; December 11, 2024)
\end{quote} 

Involuntary retirement caused by various reasons (e.g., poor health, forced unemployment) can result in income insecurity and disaster especially for low-income individuals who have yet to meet pension eligibility. In the U.K., the State Pension age is currently 66, so involuntary retirement earlier than 66 can lead to significant income gaps for retirees. The above quote from the Financial Times states that providing extra income support could help soon-to-be retirees on low incomes by filling the gaps. But what could be economic justification for doing so?

A seminal reference addressing the problem of retirement choice over the life cycle is \cite{FP}. The concept of ``optimal retirement" (i.e., individuals' decision by adjusting labor supply through the optimal choice of retirement timing) is explored and validated with the U.S. data. However, this conclusion is reached without considering disastrous income shocks and income support that are crucial elements in today's aging society when investigating investment, consumption/savings, and retirement decisions towards the end of human cycle. The presence of extra income support for low-income people can provide a rationale for their optimal retirement strategies, depending on the extent that income risk is encountered by every individual. This is the route we investigate in this paper.

We consider a representative low-income individual who exhibits the constant relative risk aversion (CRRA) utility preferences and receives the constant labor income stream. We assume the simplest possible economic setup. Specifically, the individual can trade in the financial market by allocating her wealth between one riskless bond and one risky stock. We assume the constant investment opportunity.
The individual is exposed to income disaster so that she retires involuntarily when income disaster occurs. Here, we consider extra income support provided by the government for low-income retirees who cannot meet pension eligibility yet and hence, suffer from significant income gaps.

Akin to how much to consume and invest in the stock market, retirement is one of the most important financial decisions over the life cycle. In particular, the timing of voluntary retirement is a primary concern provided that involuntary retirement occurs because of various reasons (e.g., poor health, forced unemployment). The distinct feature of this paper from the literature is that the individual must determine the optimal time to retire voluntarily prior to income disaster occurrence. We demonstrate that the decision to enter retirement in the event of income disaster is influenced, to a large extent, by the way in which how much extra income support substituting labor income is guaranteed to low-income people who are susceptible to income disaster situations. More precisely, we find a certain income support level below which they cannot afford to enter retirement due to sustainability issues and hence, the optimal decision is to delay retirement, and above which they find it achievable to retire voluntarily. This result implies that availability of the government's extra income support to cover the income gaps including social security programs or subsidies (e.g., public welfare or unemployment allowances) can be particularly important for the low-income people to attain optimal retirement in the event of income disaster.

This paper is contributing to the extant retirement literature based on Merton's portfolio choice over the life cycle \citep{Merton1969,Merton1971}. Extending \cite{KW} with optimal stopping without labor income, \cite{CS}, \cite{FP}, and \cite{CSS} have thoroughly investigated the new issues introduced by the irreversible choice of optimal retirement timing. However, these studies assumed a constant labor income stream. As far as the risk associated with labor income, \cite{DL} have explored the impact of stochastic income on the life-cycle investment, consumption/savings, and retirement choices with borrowing constraints. However, this study considered spanned income risk with the market so that individuals can diversify labor income risk by dynamically trading in the stock market. \cite{JPR} have considered forced unemployment risk in the retirement model of \cite{DL}, but the risk of forced unemployment was assumed to be hedgeable via a private unemployment insurance scheme, resulting in market completeness. \cite{BJP} have proposed an incomplete market retirement model with forced unemployment risk and \cite{JPZ} have studied the joint effect of forced unemployment risk and borrowing constraints on the life-cycle policies including retirement. \cite{CJP} have developed an analytic approach to solving optimal retirement models with disability risk in an incomplete market. However, none of these existing studies consider income disaster in the consumption/savings, investment, and retirement choices and importantly, this is the first paper to investigate the role of income support on the retirement strategies with income disaster.

The paper is organized as follows. In Section \ref{sec:2}, we describe the basic economic settings. In Section \ref{sec:3}, we propose the individual's retirement model with income disaster and investigate the role of income disaster on the retirement decisions.


\section{Basic Setup} \label{sec:2}

We first lay out the mathematical building blocks for the risk structure. The Brownian-risk in the economy is modeled by the complete probability space $(\Omega,\mathcal{F},\mathbb{P})$ on which the one-dimensional Brownian motion process $B_{t}$ later used in the stock price process is defined. The probability space $(\Omega,\mathcal{F},\mathbb{P})$ is generated by the filtration $\mathcal{F}:=\{\mathcal{F}_{t}; t\geq 0\}$, which is the usual $\mathbb{P}$-augmentation of $\sigma(B_{s}; 0\leq s\leq t)$. We denote by $\tau_{D}$ a random time of income disaster following an exponential distribution with constant intensity $\delta>0$.  For technical convenience, we assume that the Brownian-type risk and income disaster are independent. 

There are two information sets observable to an individual. First, at any time $t$ the individual knows information about past values of stock prices via the filtration $\mathcal{F}$. Second, the individual also knows information about if income disaster has taken place through the filtration $\mathcal{N}_{t}:=\sigma(\tau_{D}\wedge t)$, which is the filtration generated by the family $\tau_{D}\wedge t$, where $\tau_{D}\wedge t$ represents $\inf(\tau_{D}, t)$. We take the filtration $\mathcal{G}$ to be the smallest right continuous family of sigma-fields such that both $\mathcal{F}_{t}$ and $\mathcal{N}_{t}$ are in $\mathcal{G}_{t}$. Notice that when $\tau_{D}$ is assumed to be a $\mathcal{F}$-stopping time, the filtration $\mathcal{G}$ reduces to the filtration $\mathcal{F}$.

We consider one riskless bond and one risky stock in the financial market. The bond grows at the constant risk-free interest rate $r>0$. The risky stock price process follows a geometric Brownian motion:
$$
dS_t=\mu S_tdt+\sigma S_t dB_t,
$$
where $\mu>r$ is the expected stock return, $\sigma>0$ is the stock volatility, and $B_{t}$ is the standard one-dimensional Brownian motion process. The constant investment opportunity set is encountered by the individual.

A single consumption good (the numeraire) is considered in an infinite-horizon economy. A representative low-income individual is assumed to exhibit CRRA utility preferences. The individual has wealth $W_{t}$ at time $t$ and receives the constant labor income stream $Y_{1}>0$. The individual must determine how to allocate her wealth between the risk-free bond and the risky stock, while also deciding the optimal time to retire voluntarily. The individual receives the constant retirement benefit $Y_{2}\ge 0$. We assume that 
$$
Y_{1}>Y_{2}\ge 0.
$$ 

The individual accumulates her wealth according to the following dynamic budget constraint: $W_{0}=w$,
\begin{equation}\label{budget}
dW_{t}=\left\{
\begin{aligned}
&\left\{rW_{t}+\pi_{t}(\mu-r)-c_{t}+Y_{1}\right\}dt+\sigma\pi_{t}dB_{t},~~0\leq t<\tau,\\
&\left\{rW_{t}+\pi_{t}(\mu-r)-c_{t}+Y_{2}\right\}dt+\sigma\pi_{t}dB_{t},~~t\geq\tau,
\end{aligned}\right.
\end{equation}
where $\pi_{t}$ is the dollar amount invested in the stock market, $c_{t}$ is the amount of consumption, and $\tau$ is the optimal timing of retirement endogenously determined, which is a $\mathcal{F}$-stopping time.


\section{The Model} \label{sec:3}

The individual's model is to maximize her CRRA utility by optimally choosing consumption $c$, stock investment $\pi$, and voluntary retirement time $\tau$. The value function is defined as
$$
V(w):=\max_{(c,\pi,\tau)}\mathbb{E}\left[\int_0^\tau e^{-(\rho+\delta)t}\dfrac{c_t^{1-\gamma}}{1-\gamma}dt+e^{-(\rho+\delta)\tau}V_{D}\left(W_{\tau_{D}}\right)\right],
$$
subject to the dynamic budget constraint \eqref{budget}, where $\rho>0$ is the individual's subjective discount rate and $V_{D}(W_{\tau_{D}})$ is the post-retirement value function for low-income people who are involuntarily retired due to income disaster.  Notice that when $\tau<\infty$, the post-retirement value function directly influences the optimal pre-retirement strategy given that retirement timing $\tau$ is irreversible. The interaction of post-retirement situations with the nonlinear, option-type retirement element plays a crucial role in characterizing optimal decisions with income disaster. Our retirement model also captures the labor-leisure trade-off in a reduced-form way by assuming that people with enough wealth often retire voluntarily to enjoy their lifestyle and pursue other interest. 

Upon the occurrence of income disaster, the post-retirement value function is defined as
$$
\begin{aligned}
V_{D}(W_{\tau_{D}})&:=\max_{(c,\pi)}\mathbb{E}\left[e^{-\rho\tau_{D}}\int^{\infty}_{\tau_{D}}e^{-\rho(t-\tau_{D})}\dfrac{c^{1-\gamma}_{t}}{1-\gamma}dt\right],
\end{aligned}
$$
which is subject to the following dynamic budget constraint:
$$
dW_{t}=\left\{rW_{t}+\pi_{t}(\mu-r)-c_{t}\right\}dt+\sigma\pi_{t}dB_{t},~~t\geq\tau_{D},
$$
where the individual has no income source after income disaster because she has yet to approach mandatory pension ages to begin drawing her state pension. Following \cite{Merton1969} without considering income sources, we know the optimal relation between financial wealth and consumption as follows:
\begin{equation}\label{optimal_relation}
c_{t}=KW_{t},~~t\geq\tau_{D},
\end{equation}
where $K$ is the so-called Merton constant given by
$$
K:=r+\dfrac{\rho-r}{\gamma}+\dfrac{\gamma-1}{2\gamma^{2}}\theta^{2},\qquad\theta:=\dfrac{\mu-r}{\sigma}.
$$

The extra income support substituting labor income is provided by the government to the individual suffering from substantial income gaps until she can meet pension eligibility. The income support for low-income people continues making payments for life. We denote $0<I<Y_{1}$ by the total level of income support by public insurance and various income support schemes (e.g., government subsidies, social security programs, a Universal Basic Income, Universal Credit). The low-income people receive income support $(r+\delta)I$ per year continuously so that
$$
\int^{\infty}_{t}e^{-(r+\delta)(s-t)}(r+\delta)Ids=I,
$$
leading to the following post-disaster dynamic budget constraint:
$$
dW_{t}=\left\{rW_{t}+\pi_{t}(\mu-r)-c_{t}+(r+\delta)I\right\}dt+\sigma\pi_{t}dB_{t},~~t\geq\tau_{D}.
$$ 
With consideration such extra income sources, the post-retirement value function $V_{D}(W_{\tau_{D}})$ after the occurrence of income disaster can be obtained by following \cite{Merton1971} considering labor income as
$$
\begin{aligned}
V_{D}(W_{\tau_{D}})&=\mathbb{E}\left[e^{-\rho\tau_{D}}K^{-\gamma}\dfrac{\left(W_{\tau_{D}}+\dfrac{(r+\delta)I}{r}\right)^{1-\gamma}}{1-\gamma}\right]\\
&=\mathbb{E}\left[e^{-\rho\tau_{D}}K^{-\gamma}\dfrac{\left(K^{-1}c_{\tau_{D}}+\dfrac{(r+\delta)I}{r}\right)^{1-\gamma}}{1-\gamma}\right]\\
&=\mathbb{E}\left[e^{-\rho\tau_{D}}\dfrac{1}{K}\dfrac{\left(c_{\tau_{D}}+K\dfrac{(r+\delta)}{r}I\right)^{1-\gamma}}{1-\gamma}\right],
\end{aligned}
$$
subject to the post-retirement dynamic budget constraint in \eqref{budget}, where the second equality results from the optimal relation \eqref{optimal_relation} between financial wealth and consumption. By integrating out the exponential distribution of income disaster with intensity $\delta$, we now define the post-retirement value function just before income disaster as
$$
V_{D}(w):=\max_{(c,\pi)}\mathbb{E}\left[\int^{\infty}_{0}e^{-(\rho+\delta)t}\dfrac{\delta}{K}\dfrac{\left(c_{t}+K\dfrac{(r+\delta)}{r}I\right)^{1-\gamma}}{1-\gamma}dt\right],
$$
subject to the post-retirement dynamic budget constraint in \eqref{budget}, where the subjective discount rate now incorporates the constant income disaster intensity. Without loss of generality, we simplify the post-retirement value function $V_{D}(w)$ with $L:=K\frac{(r+\delta)}{r}I$ for notational convenience by setting that $\delta=K$ so that
\begin{equation}\label{post_disaster_value}
V_{D}(w)=\max_{(c,\pi)}\mathbb{E}\left[\int^{\infty}_{0}e^{-(\rho+\delta)t}\dfrac{(c_{t}+L)^{1-\gamma}}{1-\gamma}dt\right],
\end{equation}
subject to the post-retirement dynamic budget constraint in \eqref{budget}. Here, $L$ can be thought of as the level of living standard the individual wishes to maintain, where social security benefits can impute the living standard. Notice that the subjective discount rate for discounting future consumption utility values is $\rho+\delta$ incorporating a constant income disaster intensity, $\delta$. This reflects the fact that, in the event of income disaster, future consumption price becomes more expensive, leading the low-income individual to be in a high marginal utility situation so that she is willing to give up more consumption now than without income disaster for financing more expensive future consumption costs.

By the principle of dynamic programming, we finally derive the following model:
\begin{equation}\label{main model}
V(w)=\max_{(c,\pi,\tau)}\mathbb{E}\left[\int_0^\tau e^{-(\rho+\delta)t}\dfrac{c_t^{1-\gamma}}{1-\gamma}dt+e^{-(\rho+\delta)\tau}\int^{\infty}_{\tau}e^{-(\rho+\delta)(t-\tau)}\dfrac{(c_{t}+L)^{1-\gamma}}{1-\gamma}dt\right],
\end{equation}
subject to the dynamic budget constraint \eqref{budget}. By solving the model, we provide a main theorem regarding the low-income individual's optimal retirement strategies with income disaster and income support.

\begin{thm}
We find a certain income support level below which the low-income people cannot afford to enter retirement and hence, the optimal decision is to delay retirement. More specifically, when the income support level $L$ is less than or equal to $Y_{1}-Y_{2}$ ($L\le Y_{1}-Y_{2}$), which we refer to as income gaps, the optimal time to retire satisfies $\tau=\infty$.

When the income support level $L$ is greater than the income gaps $Y_{1}-Y_{2}$ ($L> Y_{1}-Y_{2}$), the low-income people can achieve optimal retirement with the optimal time to retire satisfying $0<\tau<\infty$, which is determined by the wealth threshold for retirement as follows:
$$
\tau=\inf\{t\geq 0:W_{t}\geq\overline{w}\},
$$
where $\overline{w}$ is the retirement wealth threshold.
\end{thm}
\begin{proof}
We prove the theorem in two steps. We will apply the duality approach of \cite{KW} with optimal stopping to solve the model \eqref{main model} by introducing the standard state price density as follows:
$$
H_{t}:=e^{-\left(r+\frac{1}{2}\theta^{2}\right)t-\theta B_{t}},
$$
implying a dynamically complete market. 

\textit{Step 1 (Post-Retirement Value Function).} We first solve the post-retirement value function $V_{D}(w)$ in \eqref{post_disaster_value} just before income disaster by applying the duality approach. The post-retirement budget constraint in \eqref{budget} can be converted into the following static budget constraint:
$$
\mathbb{E}\left[\int^{\infty}_{0}H_{t}c_{t}dt\right]\le w+\dfrac{Y_{2}}{r}.
$$
Following the standard Lagrangian approach, we can obtain the following inequality: for the Lagrange multiplier $z>0$,
$$
\begin{aligned}
V_{D}(w)-wz&\leq\max_{(c,\pi)}\left[\mathbb{E}\left[\int^{\infty}_{0}e^{-(\rho+\delta)t}\dfrac{(c_{t}+L)^{1-\gamma}}{1-\gamma}dt\right]-z\,\mathbb{E}\left[\int^{\infty}_{0}H_{t}c_{t}dt\right]+\dfrac{Y_{2}}{r}z\right]\\
&=\max_{(c,\pi)}\mathbb{E}\left[\int^{\infty}_{0}e^{-(\rho+\delta)t}\left\{\dfrac{(c_{t}+L)^{1-\gamma}}{1-\gamma}-ze^{(\rho+\delta)t}H_{t}c_{t}\right\}dt\right]+\dfrac{Y_{2}}{r}z.
\end{aligned}
$$
We then introduce the following dual variable:
$$
z_{t}:=ze^{(\rho+\delta)t}H_{t}
$$
implying the stochastic dynamics as follows:
$$
dz_{t}=z_{t}\left\{\left(\rho+\delta-r\right)dt-\theta dB_{t}\right\},~~z_0=z>0.
$$
We construct the dual value function as follows:
$$
\begin{aligned}
\tilde{V}_{D}(z)&:=\mathbb{E}\left[\int^{\infty}_{0}e^{-(\rho+\delta)t}\max_{c\geq 0}\left\{\dfrac{(c_{t}+L)^{1-\gamma}}{1-\gamma}-c_{t}z_{t}\right\}dt\right]+\dfrac{Y_{2}}{r}z\\
&=\mathbb{E}\left[\int^{\infty}_{0}e^{-(\rho+\delta)t}\left\{\left(\dfrac{\gamma}{1-\gamma}z^{\frac{\gamma-1}{\gamma}}_{t}+L z_{t}\right)\textbf{1}_{\{0<z_{t}<L^{-\gamma}\}}+\dfrac{1}{1-\gamma}L^{1-\gamma}\textbf{1}_{\{z_{t}\geq L^{-\gamma}\}}\right\}dt\right]+\dfrac{Y_{2}}{r}z,
\end{aligned}
$$
where $\textbf{1}$ is the indicator function. The Feynman-Kac formula leads to the ordinary differential equation (ODE):
\begin{equation}
\begin{aligned} \label{eq:ODE1}
\dfrac{1}{2}\theta^{2}z^{2}\tilde{v}_{D}''(z)&+(\rho+\delta-r)z\tilde{v}_{D}'(z)-(\rho+\delta)\tilde{v}_{D}(z)\\
&+\Big(\dfrac{\gamma}{1-\gamma}z^{\frac{\gamma-1}{\gamma}}+L z\Big)\textbf{1}_{\{0<z<L^{-\gamma}\}}+\dfrac{1}{1-\gamma}L^{1-\gamma}\textbf{1}_{\{z\geq L^{-\gamma}\}}=0,
\end{aligned}
\end{equation}
where $\tilde{v}_{D}(z):=\tilde{V}_{D}(z)-(Y_{2}z)/r$. By solving the ODE \eqref{eq:ODE1}, we now obtain the post-retirement dual value function $\tilde{V}_{D}(w)$ as
$$
\tilde{V}_{D}(z)=\dfrac{Y_{2}}{r}z+\left\{
\begin{aligned}
&Az^{m_{+}}+\dfrac{\gamma}{K(1-\gamma)}z^{\frac{\gamma-1}{\gamma}}+\dfrac{L}{r}z,~~\mbox{if}~~0<z<L^{-\gamma},\\
&Bz^{m_{-}}+\dfrac{1}{(\rho+\delta)(1-\gamma)}L^{1-\gamma},~~~\mbox{if}~~z\geq L^{-\gamma},
\end{aligned}\right.
$$
where
$$
\begin{aligned}
&A=-\dfrac{2L^{1+\gamma(m_{+}-1)}}{(m_{+}-m_{-})m_{+}(m_{+}-1)\{\gamma(m_{+}-1)+1\}\theta^{2}}<0,\\
&B=-\dfrac{2L^{1+\gamma(m_{-}-1)}}{(m_{+}-m_{-})m_{-}(m_{-}-1)\{\gamma(m_{-}-1)+1\}\theta^{2}}>0,
\end{aligned}
$$
and $m_{+}>1$ and $m_{-}<\min \big\{ 0, 1 - 1/\gamma \big\}$ are the two roots of the following quadratic equation:
$$
\dfrac{1}{2}\theta^{2}m^{2}+\left(\rho+\delta-r-\dfrac{1}{2}\theta^{2}\right)m-(\rho+\delta)=0.
$$

\textit{Step 2 (Pre-Retirement Value Function).} The dynamic budget constraint \eqref{budget} can be converted into the following static budget constraint:
$$
\mathbb{E}\left[\int^{\tau}_{0}H_{t}c_{t}dt+H_{\tau}\left(W_{\tau}+\dfrac{Y_{1}}{r}\right)\right]\leq w+\dfrac{Y_{1}}{r}.
$$
As a result, we obtain the following inequality: for the Lagrange multiplier $z>0$,
$$
\begin{aligned}
V(w)-wz&\leq\max_{(c,\pi,\tau)}\mathbb{E}\left[\int^{\tau}_{0}e^{-(\rho+\delta)t}\left(\dfrac{c^{1-\gamma}_{t}}{1-\gamma}-ze^{(\rho+\delta)t}H_{t}c_{t}\right)dt\right.\\
&\qquad\left.+e^{-(\rho+\delta)\tau}\left\{V_{D}(W_{\tau})-ze^{(\rho+\delta)\tau}H_{\tau}\left(W_{\tau}+\dfrac{Y_{1}}{r}\right)\right\}\right]+\dfrac{Y_{1}}{r}z.
\end{aligned}
$$
We then establish the dual value function as follows: 
$$
\begin{aligned}
\tilde{V}(z)&:=\max_{\tau}\,\mathbb{E}\left[\int^{\tau}_{0}e^{-(\rho+\delta)t}\max_{c\geq 0}\left(\dfrac{c^{1-\gamma}_{t}}{1-\gamma}-c_{t}z_{t}\right)dt\right.\\
&\qquad\left.+e^{-(\rho+\delta)\tau}\left\{V_{D}(W_{\tau})-ze^{(\rho+\delta)\tau}H_{\tau}\left(W_{\tau}+\dfrac{Y_{1}}{r}\right)\right\}\right]+\dfrac{Y_{1}}{r}z\\
&=\max_{\tau}\,\mathbb{E}\left[\int^{\tau}_{0}e^{-(\rho+\delta)t}\dfrac{\gamma}{1-\gamma}z^{\frac{\gamma-1}{\gamma}}_{t}dt+e^{-(\rho+\delta)\tau}\left\{\tilde{V}_{D}(z_{\tau})-\dfrac{Y_{1}}{r}z_{\tau}\right\}\right]+\dfrac{Y_{1}}{r}z.
\end{aligned}
$$
The dual value $\tilde{V}$ can be found by solving the variational inequality as follows \citep{BL}:
$$
\left\{
\begin{aligned}
&\dfrac{1}{2}\theta^{2}z^{2}\tilde{\Psi}''(z)+(\rho+\delta-r)z\tilde{\Psi}'(z)-(\rho+\delta)\tilde{\Psi}(z)\\
&~+\dfrac{\gamma}{1-\gamma}z^{\frac{\gamma-1}{\gamma}}-\left(\dfrac{\gamma}{1-\gamma}z^{\frac{\gamma-1}{\gamma}}+L z\right)\textbf{1}_{\{0<z<L^{-\gamma}\}}\\
&~-\dfrac{1}{1-\gamma}L^{1-\gamma}\textbf{1}_{\{z\geq L^{-\gamma}\}}+(Y_{1}-Y_{2})z=0,~~z\ge\overline{z},\\
&\dfrac{1}{2}\theta^{2}z^{2}\tilde{\Psi}''(z)+(\rho+\delta-r)z\tilde{\Psi}'(z)-(\rho+\delta)\tilde{\Psi}(z)\\
&~+\dfrac{\gamma}{1-\gamma}z^{\frac{\gamma-1}{\gamma}}-\left(\dfrac{\gamma}{1-\gamma}z^{\frac{\gamma-1}{\gamma}}+L z\right)\textbf{1}_{\{0<z<L^{-\gamma}\}}\\
&~-\dfrac{1}{1-\gamma}L^{1-\gamma}\textbf{1}_{\{z\geq L^{-\gamma}\}}+(Y_{1}-Y_{2})z< 0,~~0<z<\overline{z},\\
&\tilde{\Psi}(z)< 0,~~z>\overline{z},\\
&\tilde{\Psi}(z)=0,~~0<z\le\overline{z},
\end{aligned}\right.
$$
where $\tilde{\Psi}(z):=\tilde{V}(z)-\tilde{V}_{D}(z)$ and $\overline{z}>0$ is a dual threshold of the dual variable $z$ below which it is optimal to enter retirement. The variational inequality has the following stochastic representation:
\begin{equation}\label{stochastic_representation}
\begin{aligned}
\tilde{\Psi}(z)=\max_{\tau}\,\mathbb{E}\Bigg[\int^{\tau}_{0}e^{-(\rho+\delta)t} g(z_{t}) dt\Bigg]
\end{aligned}
\end{equation}
with the optimal time to retire $\tau$ determined by the dual threshold $\overline{z}$ as
$$
\tau=\inf\{t\geq 0: z_{t}\leq\overline{z}\}
$$
and
$$
g(z) := \dfrac{\gamma}{1-\gamma}z^{\frac{\gamma-1}{\gamma}}-\left(\dfrac{\gamma}{1-\gamma}z^{\frac{\gamma-1}{\gamma}}+L z\right)\textbf{1}_{\{0<z<L^{-\gamma}\}}-\dfrac{1}{1-\gamma}L^{1-\gamma}\textbf{1}_{\{z\geq L^{-\gamma}\}}+(Y_{1}-Y_{2})z.
$$
Variation of parameters leads to 
\begin{align}
  \tilde{\Psi}(z) = \begin{cases}
   - \frac{2 z^{m_{-}}}{\theta^{2} (m_{+} - m_{-})} \displaystyle\int_{0}^{\overline{z}} y^{- 1 - m_{-}} g(y) d y \\
   \qquad + \frac{2}{\theta^{2} (m_{+} - m_{-})} \left( z^{m_{+}}  \displaystyle\int_{z}^{\infty} y^{- 1 - m_{+}} g(y) d y + z^{m_{-}} \int_{0}^{z} y^{- 1 - m_{-}} g(y) d y \right), \quad & ~ z > \overline{z}, \\
   0, & ~ 0 < z \leq \overline{z}
   \end{cases} \label{Tilde_z}
\end{align}
with 
\begin{align}
  \int_{\overline{z}}^{\infty} y^{- 1 - m_{+}} g(y) d y = 0. \label{Equation_overline_z}
\end{align}
We observe the following relations:
\begin{equation} \label{eq:relation}
g(z)= \begin{cases}
(Y_{1}-Y_{2}-L)z,&~~0<z< L^{-\gamma},\\
\dfrac{\gamma}{1-\gamma}z^{\frac{\gamma-1}{\gamma}}-\dfrac{1}{1-\gamma}L^{1-\gamma}+(Y_{1}-Y_{2})z,&~~z\ge L^{-\gamma}.
\end{cases}
\end{equation}
Observe that $z \mapsto g(z)$ is continuously differentiable, convex on the entire domain, and especially strictly convex on $z\ge L^{-\gamma}$.

When the income support level $L$ is less than $Y_{1}-Y_{2}$, the relations \eqref{eq:relation} and the properties of $g$ therefore show that the integrand of the stochastic representation \eqref{stochastic_representation} is strictly positive and hence, the optimal time to retire must be $\tau=\infty$.

When the income support level $L$ is equal to $Y_{1}-Y_{2}$, we get $g(z)=0$ on $0<z<L^{-\gamma}$ and $g(z)>0$ on $z\ge L^{-\gamma}$. Since the event $\{ z_{t}\ge L^{-\gamma} \}$ has a positive probability for each $t > 0$, the optimal time to retire must be $\tau=\infty$ to achieve \eqref{stochastic_representation}.

When the income support level $L$ is greater than $Y_{1}-Y_{2}$, we have $g(z)<0$ on $0<z<L^{-\gamma}$. The properties of $g$ conclude that there exists a unique $j \in (L^{- \gamma}, \infty)$ such that 
\begin{align}
g(z)  \begin{cases}
   < 0 \qquad \text{if } z < j, \\
   = 0 \qquad \text{if } z = j, \\
   > 0 \qquad \text{if } z > j.
  \end{cases}
\end{align}
Notice that 
$$
\int_{0}^{\infty} y^{- 1 - m_{+}} g(y) d y < 0 < \int_{j}^{\infty} y^{- 1 - m_{+}} g(y) d y.
$$ Since $z \mapsto \int_{z}^{\infty} y^{- 1 - m_{+}} g(y) d y$ is continuous, the intermediate value theorem suggests that there exists $\overline{z} \in (0, j)$ such that $\overline{z}$ solves \eqref{Equation_overline_z}. Uniqueness is shown by the fact that $z \mapsto \int_{z}^{\infty} y^{- 1 - m_{+}} g(y) d y$ is strictly increasing in $(0, j)$. Therefore, the optimal time to retire satisfies $0<\tau<\infty$.

By the duality, we introduce $\overline{w}:=-\tilde{V}'(\overline{z})$, which we refer to as the retirement wealth threshold so that the optimal time to retire $\tau$ can be characterized as 
$$
\tau=\inf\left\{t\geq 0: W_{t}\geq\overline{w}\right\},
$$
which now completes the proof of the theorem.
\end{proof}


The presence of income disaster introduces a new friction into the economy that affects both investment and consumption/savings choices, and hence plays a crucial role in the interactions among consumption/savings, investment, and retirement choices for individuals. The effects of income disaster on the low-income people's financial sustainability are two-fold. On the one hand, when income disaster occurs under various income disaster scenarios (e.g., job loss, accident, permanent disability), labor income $Y_{1}$ drops to zero permanently. On the other hand, the low-income people are not entitled to pension benefits $Y_{2}$ yet as a result of forced unemployment before mandatory pension ages. Here, income gaps arise measured by the difference $Y_{1}-Y_{2}$ between labor income $Y_{1}$ and retirement benefits $Y_{2}$. 

Similar to how much to consume and invest in the stock market, retirement is one of the most important financial decisions over the life cycle. In particular, the timing of voluntary retirement is a primary concern given that involuntary retirement can be caused due to various reasons (e.g., poor health, forced unemployment). Our theorem suggests that if social security programs or subsidies (e.g., public welfare or unemployment allowances) cannot cover the income gaps, the optimal decision is to delay retirement. Without consideration of enough income support, the low-income people's ability to recover from and respond to income disaster is further impeded by their ill-preparedness for future consumption needs so that they cannot afford to enter retirement. In other words, availability of the government's extra income support can be particularly important for the low-income people to achieve optimal retirement in the event of income disaster. The decision to retire is likely determined by levels of wealth, so the recovery of income in the aftermath of income disaster is crucial for asset accumulation purposes towards the retirement wealth threshold. The explicit efforts by governments should, therefore, focus on greater access to a wide range of government safety nets for the low-income people to recover from and increase their resilience against income disaster.



\end{document}